# Local volume conservation in concentrated electrolytes is governing charge transport in electric fields

*Martin Lorenz,[a] Franziska Kilchert,[b] Pinchas Nürnberg,[a,†] Max Schammer,[b] Arnulf Latz,[b,c,d] Birger Horstmann,[b,c,d] Monika Schönhoff*[a]*

[a] University of Münster, Institute of Physical Chemistry, Corrensstrasse 28/30, 48149 Münster, Germany

[b] German Aerospace Center, Pfaffenwaldring 38-40, 70569 Stuttgart, Germany

[c] Helmholtz Institute Ulm, Helmholtzstraße 11, 89081 Ulm, Germany

[d] Universität Ulm, Albert-Einstein-Allee 47, 89081 Ulm, Germany

*Email: schoenho@uni-muenster.de, phone +49-251-8323419

[†] present address: Pinchas Nürnberg, KIT, Institute of Meteorology and Climate Research - Atmospheric Environmental Research (IMK-IFU), Kreuzeckbahnstraße 19, 82467 Garmisch-Partenkirchen, Germany

**ABSTRACT**

While ion transport processes in concentrated electrolytes, e.g. based on ionic liquids (IL), are a subject of intense research, the role of conservation laws and reference frames is still a matter of debate. Employing electrophoretic NMR, we show that momentum conservation, a typical prerequisite in molecular dynamics (MD) simulations, is not governing ion transport. Involving density measurements to determine molar volumes of distinct ion species, we propose that conservation of local molar species volumes is the governing constraint for ion transport. The experimentally quantified net volume flux is found as zero, implying a non-zero local momentum flux, as tested in pure ILs and IL-based electrolytes for a broad variety of concentrations and chemical compositions. This constraint is consistent with incompressibility, but not with a local application of momentum conservation. The constraint affects the calculation of transference numbers as well as comparisons of MD results to experimental findings.

**TOC GRAPHICS**

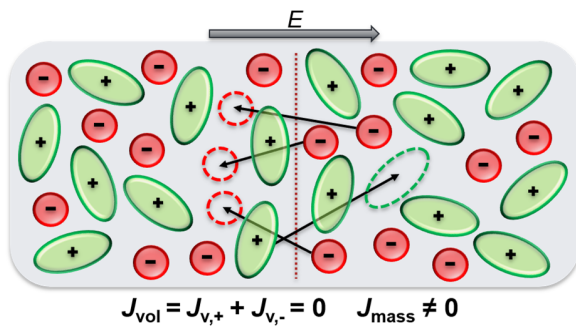

Research efforts in the field of lithium conducting electrolytes are strongly driven by the increasing demand regarding energy storage devices.[1] One focus is the substitution of widely used carbonate-based liquid electrolytes by less flammable systems, while maintaining advantageous transport properties.[2, 3] Highly concentrated electrolytes are an alternative, as the reduced vapor pressure significantly improves safety and enables high-temperature applications.[4-6] In particular, ionic liquid (IL)- based systems consisting purely of ions form highly concentrated electrolytes with beneficial properties.[7, 8] However, the transport properties in concentrated ionic systems are dominated by strong Coulomb interaction, rendering them far more complex as compared to dilute electrolytes. Ion-ion correlations have to be taken into account, which include correlated transport between different species as well as distinct ion correlations between ions of the same species.[9-11]

For experimental characterization of ion transport Pulsed Field Gradient (PFG)-NMR is a widely applied and beneficial tool, as it provides species-selective self-diffusion coefficients.[12-14] However, it does neither deliver directional transport information, nor can it distinguish transport contributions of charged and uncharged species, respectively. In contrast, electrophoretic NMR (eNMR) yields drift velocities and directions of charged species in an electric field.[15-17] Initially only established in dilute solutions, the method was lately optimized to include measures to suppress convective flow induced by Joule heating in highly conductive samples.[18-20] This finally allowed application of eNMR to concentrated electrolytes such as ionic liquids.[21, 22] This technique already led to a deeper understanding of transport processes, an important finding being the fact that Li transference numbers in Li salt in IL electrolytes are negative, as directly evidenced from negative lithium mobilities. This is interpreted as a vehicular transport mechanism of lithium in net negatively charged Li-anion clusters rather than $Li^+$ hopping between adjacent coordination environments,[23] and demonstrated to be a general feature of IL-based electrolytes.[24] ENMR experiments could further demonstrate how this mechanism can be overcome by chelating additives, leading to positive Li transference numbers.[25, 26] In addition to ionic species, the drift motions of uncharged species, for example when coordinating to Li$^+$, can be identified, clarifying their role in Li$^+$ transport.[27] Even beyond species-based models of transport, electrophoretic mobilities can contribute to quatify ion correlations, i.e. Onsager coefficients,[11] or explain the role of ion correlations in superionicity of liquid electrolytes.[28] In parallel, however, critical discussions arose around the technique, centered around the sign of transference numbers as well as relevant reference frames. On the one hand, there was a controversy about the sign of the Li$^+$ transference number in salt in poly(ethylene oxide), which resulted as positive from the Bruce-Vincent method,[29] as well as from eNMR,[30, 31] but appeared to be negative in a certain concentration regime when determined from electrochemical experiments based on concentrated solution theory.[32, 33] This controversy was only recently resolved by realizing a fundamental difference of the reference frames relevant to different methods. Accordingly, concentrated solution theory-based data could be reconciled with MD results by a transformation from their inherent solvent-fixed frame to the center of mass (COM) frame employed in MD simulations.[34] In IL-based liquid electrolytes on the other hand, recent discussions dealt with suitable reference frames and their relevance for the transference number.[35, 36] Mobilities determined by eNMR are usually displayed as they are measured, i.e. in the laboratory reference frame. While for a closed system the basic physical principle of momentum conservation implies a zero net mass flux in the electrolyte, eNMR measurements on highly concentrated electrolytes show a COM drift towards the anode.[36] Even in pure ILs, where only two oppositely charged ionic species are present, the principle of momentum conservation is locally violated and a negative COM drift occurs. Although in principle the COM drift can be explained by the non-closed configuration of the setup (momentum can be transferred to the sample walls and the electrodes), it led to an intense discussion about the appropriate choice of reference frame.[35, 36] In this context, different proposals were discussed, such as using the mobility of one of the ionic species as reference for the others,[35] or transforming drift velocities to the COM frame where local momentum conservation applies.[36] The latter transformation was also employed to achieve compatibility with Onsager theory of ion correlations and to extract Onsager coefficients in solvate ionic liquids.[11] Similarly, MD simulations intrinsically assume a fixed COM, incompatible with the observation of the COM drift in the open system of the experiment.

The above controversies give rise to the very fundamental question of an underlying constraint governing transport processes in concentrated electrolytes, as mobilities of different ion species are not independent quantities, since they are connected by the overall conductivity of the sample. In theoretical treatments of concentrated aqueous solutions the importance of volume effects due to finite partial molar volumes for interfacial and bulk properties is well known.[37-39] Concentrated electrolytes form multicomponent systems, where the individual species partial molar volumes can affect interfacial properties[40, 41] as well as drift velocities.[42, 43] Nevertheless, standard models often overlook its relevance. Furthermore, confusion may arise as different definitions of the drift velocity are applied in modelling. These include the solvent species as reference,[44, 45] the center-of-mass drift,[40, 46] or the volume-averaged drift velocity.[47, 48]

In this letter we prove experimentally that charge transport in pure ionic liquids and respective Li salt containing electrolytes is governed by their incompressibility, inducing an external constraint. Accordingly, the molar volume of each molecular species is conserved and the local net volume flux in the system is zero. This hypothesis is tested for a broad range of ILs and ternary electrolytes (Li salt in IL). We determine the molar volumes of all ion species by density measurements and calculate the net volume flux as well as the net mass flux under an electric field employing ionic mobilities determined by eNMR. The experimental results are consistent with a zero local volume flux, but not with a zero local mass flux, thus proving that local molar volume conservation is the constraint governing the relation of mobilities of different species in concentrated electrolyte systems, rather than local momentum conservation. This applies to open systems in experiments with electrolyte between two electrodes, and it differs fundamentally from MD simulations in closed systems with intrinsic assumption of momentum conservation.

First of all, it is important to realize the conditions of the eNMR experiment to detect ion drift velocities in the bulk. **Figure 1b** shows a sketch of the conditions in the measurement cell, indicating the active volume (dashed red line) covered by the RF coil. The volume up to ca. 1 mm distance from the electrode is not covered by the active volume. Thus, the quantities determined by NMR represent ensemble averages over the active volume from which the NMR signal is taken. We assume a plug flow of either ion species with an ion-specific drift velocity $u_i$, defined in the laboratory frame. During an electric field pulse, $u_i$ is constant in time and in space. Electric field pulses of a duration of typically 100 ms lead to a displacement of about 0.5 μm under an electric field of $E$ = 50 V/cm. In the eNMR pulse sequence (see **Figure 1a**) the sign of the voltage pulses is inverted, resulting in a shuttling of the ions between positions z and z+Δz by a small Δz ~ 0.5 μm. For all following considerations, a constant concentration of each ion species $i$ in the active volume is assumed. Electrode processes are not considered, as any molecules in the interfacial region are not detected. Note that due to the short pulse lengths and displacements the conditions are fundamentally different from electrochemical experiments involving polarization.

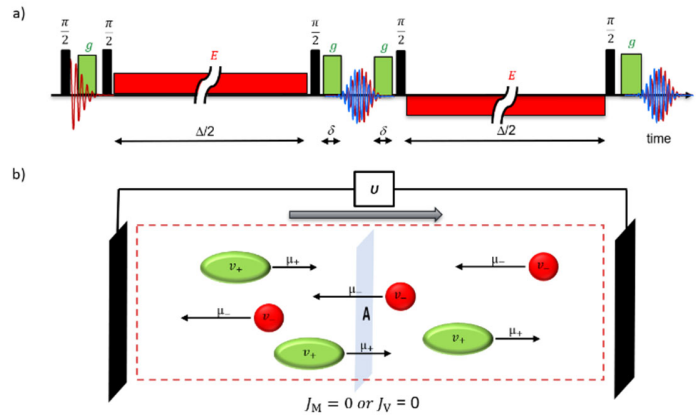

**Figure 1.** a) Double stimulated echo pulse sequence[49] employed in eNMR experiments with electric field pulses of alternating sign ($E$), gradient pulses $g$ and radiofrequency pulses ($\pi/2$) b) Sketch of ion drift during an electric field pulse, showing electrodes (black) and the active volume (dashed red line). The latter is the volume covered by the RF coil, from which the signal is taken, while electrode processes are not detected. Electric field pulses are applied with alternating sign and shuttle ions by sub-μm distances in subsequent pulses, thus the concentration of anions (red) and cations (green) in the active volume is assumed to be constant. The net volume flux $J_V$ through a cross-sectional area A perpendicular to the electric field is indicated.

We consider an electrolyte under steady state conditions during an electric field pulse. In case of validity of local momentum conservation the total mass flux through the area A, see **Figure 1b**, $J_M$, expressed as a sum over all species $i$, would be zero, i.e.

$$J_M = \sum_i N_i \cdot M_i \cdot u_i = 0 \qquad (1)$$

where $M_i$ is the molar mass and $N_i$ the number density. We note here that we distinguish between *global* momentum conservation as a fundamental physical principle, applying in closed systems, and *local* momentum conservation based on local mass fluxes as described in eq (1).

For the simple case of a pure ionic liquid (IL) with only two species indicated by + and -, respectively, and equal number densities $N_+ = N_-$ eq (1) simplifies to

$$\frac{\mu_+}{|\mu_-|} = \frac{u_+}{|u_-|} = \frac{M_-}{M_+}. \qquad (2)$$

Here, $\mu_+$ and $\mu_-$ are the cation and anion mobility, respectively. The plot in **Figure 2a** gives the mobility ratio, plotted against the mass ratio, for different ILs as determined by eNMR. Open symbols are taken from an earlier publication,[22] while full symbols represent new data of the present work. The supplementary information provides exemplary phase shift data and their fit by eq (S1) in **Figure S1**, as well as a table with all newly measured mobility values, **Table S2**.

Generally, the mobility ratios in **Figure 2a** are below the diagonal, the latter representing the case of validity of local momentum conservation. Obviously, local momentum conservation is not valid in the eNMR experiment. In fact, the electrolyte forms an open system, being in contact with the electrodes and cell walls, accordingly, momentum can be transferred to electrodes and cell walls. In addition, momentum is introduced by the electric field exerting a force on the ions. An interesting finding is that the ratio $\mu_+/\mu_-$ seems to be biased in the direction of the anion, implying that momentum transfer to the anode rather than to the cathode dominates the deviation.

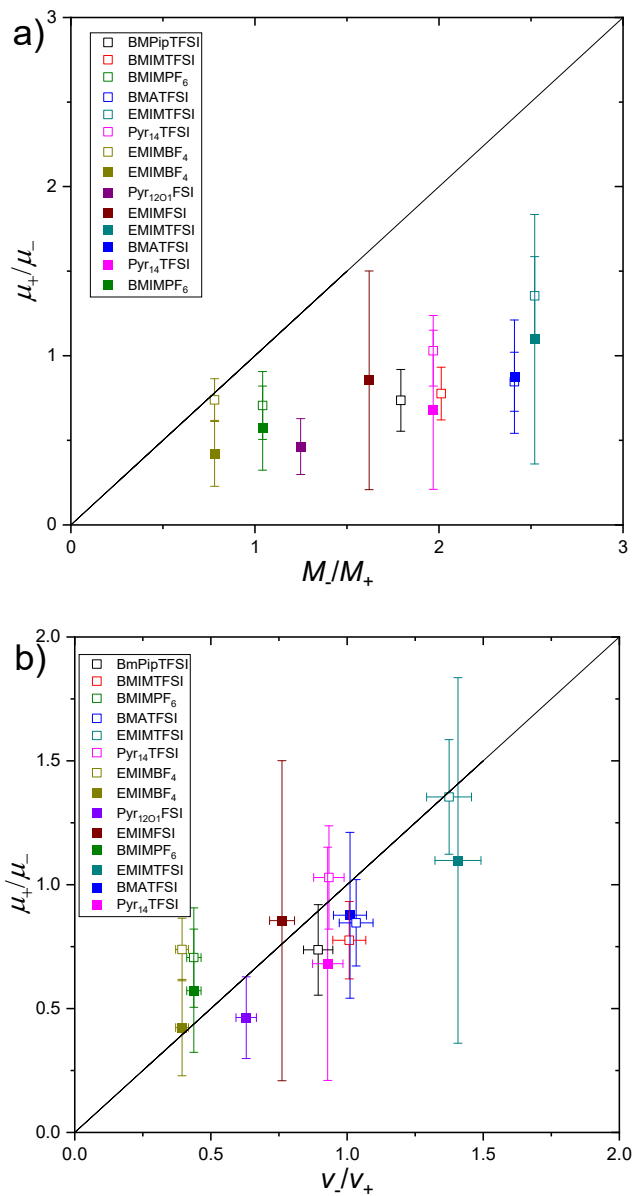

**Figure 2.** Mobility ratio a) vs. mass ratio and b) vs. volume ratio for different ionic liquids. The line represents the diagonal. Mobility data partly taken from earlier work[22, 36] (open symbols) and investigated in this work (solid symbols). All molar volumes taken at 295 K, except full symbols for EMIMTFSI, BMATFSI und Pyr$_{14}$TFSI at 300 K.

With local momentum conservation being invalid, the question arises which boundary condition is governing the relative ion fluxes in the open system. Our hypothesis is that instead of local momentum conservation, conservation of the local molar volume is the relevant boundary condition, which controls the ratio of cation and anion flux. This hypothesis is motivated by the known low compressibility of concentrated electrolytes such as ILs.[50-52] In the following, we will test this hypothesis.

A net volume flux towards one of the electrodes would lead to a volume accumulation during an electric field pulse. We assume that this is prevented by incompressibility and the net volume flux through an area A, which is perpendicular to the electric field (see **Figure 1b**) is zero. In this case, eq (3) replaces eq (1) and it is

$$J_V = \sum_i N_i \cdot v_i \cdot u_i = 0 \qquad (3)$$

with the partial molar volumes $v_i$, here $v_+$ and $v_-$ of the individual ionic species. In the case of a simple IL with two species it is

$$\frac{\mu_+}{|\mu_-|} = \frac{u_+}{|u_-|} = \frac{v_-}{v_+}. \qquad (4)$$

This expression bears a formal analogy to eq (2), with molar mass being replaced by molar volume, representing a local volume constraint instead of local mass flux (momentum) constraint. Note that in a compressible fluid the net volume flux might not be zero, but dependent on the compressibility.

In order to test the validity of eq (4), the molar volumes of all ionic species involved in the different ILs are required. Their calculation, derived from density measurements with varying Li salt concentration is presented in the SI, section 4. It is based on a known value of the $Li^+$ ion volume, calculated from an ion radius of $r_{Li^+} = 76$ pm as the effective ionic radius of a sixfold coordinated lithium ion.[53] The resulting species molar volumes are independent of the composition of the electrolyte concerning varying IL components or varying Li concentration. Employing the resulting molar volumes (given in Table (S3)), a test of eq (4) is shown in **Figure 2b**. Here, the data points follow the ideal behavior given by the diagonal, with no systematic deviation, validating the assumption of constant molar volume.

In addition to this representation, the calculation of the total molar volume flux $J_V$ is another appropriate method to test this concept. In order to judge the validity of eq (3), $J_V$ can be related to the sum of the moduli of the volume fluxes of the individual species $J_{V,tot}$:

$$\frac{J_V}{J_{V,tot}} = \frac{J_V}{|J_+|+|J_-|} = \frac{\mu_+ \cdot v_+ + \mu_- \cdot v_-}{|\mu_+ \cdot v_+| + |\mu_- \cdot v_-|} \qquad (5)$$

Note that $\mu_-$ is negative due to the negative drift direction. The corresponding values for all investigated ILs are shown in **Figure S2**. In accordance with the results of **Figure 2b**, $\frac{J_V}{J_{V,tot}}$ is zero within error for most data points. Slight deviations occur in case of previously published data for the anions $BF_4^-$ and $PF_6^-$, which will be discussed further below.

In conclusion, the concept of a local molar volume constraint provides a significantly more valid description of the transport processes occurring in purely ionic systems than the concept of local momentum conservation. This is in striking contrast to MD simulations, where a closed system is assumed and local momentum conservation is an a priori assumption.

Next, we are probing the relevance of the volume conservation constraint for systems of higher complexity incorporating more than two ion species, namely ternary lithium salt in IL systems. Employing a common anion of the IL and the Li salt yields systems with three distinct ion species. Here, a simple relationship between mobilities and molar mass or molar volumes, such as eq (2) or (4) cannot be established. However, the calculation of a relative molar volume flux in analogy to eq (5) serves as an alternative. Considering cation (+), anion (-) and lithium ion (Li) in Li salt in IL systems with the same anion, the mass flux $J_M$ and the volume flux $J_V$ of all three species can again be related to the sum of the moduli of all fluxes, yielding

$$\frac{J_{M,V}}{J_{M,V,tot}} = \frac{\sum_i J_{M,V,i}}{\sum_i |J_{M,V,i}|}. \qquad (6)$$

For testing the validity of either approach, we use mobility data already published by Brinkkötter *et al.* regarding $Pyr_{12O1}FTFSI/LiFTFSI$-electrolytes[54] and Gouverneur *et al.* concerning $EMIMBF_4/LiBF_4$- and $EMIMTFSI/LiTFSI$-systems.[23] To broaden the data base, we measure eNMR mobilities of FSI-, TFSI- and $BF_4$-based electrolytes, including consideration of previously investigated systems at different salt concentrations. Exemplary eNMR phase shift data and their evaluation by a linear fit according to eq (S1) are given in **Figure S5**. The resulting mobility values of the newly measured systems are tabulated in **Table S7**.

Partial molar volumes $v_i$ required to calculate the net volume flux $J_V$ and the total volume flux $J_{V,tot}$ for the FSI, TFSI and BF$_4$-based electrolytes are used as calculated above. The asymmetric anion FTFSI contains one structural element of FSI and TFSI each, thus $v_{FTFSI}$ for the Pyr$_{12O1}$FTFSI-based systems is estimated as the average of $v_{FSI}$ and $v_{TFSI}$ calculated for their respective Pyr$_{12O1}$-ILs at 295 K.

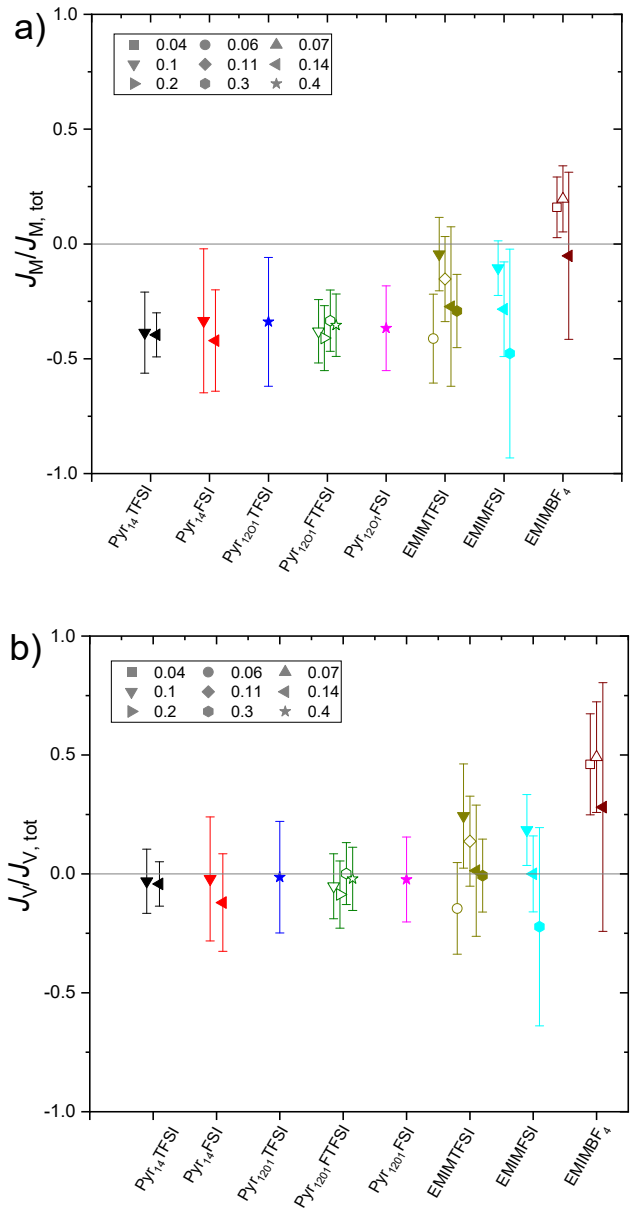

**Figure 3.** Comparison of $\frac{J_M}{J_{M,tot}}$ (a) and $\frac{J_V}{J_{V,tot}}$ (b) calculated from mobilities of FSI-, TFSI- and BF$_4$-based electrolytes (this work, full symbols) and from previously published mobilities of other systems (open symbols, eNMR data of Pyr$_{12O1}$FTFSI/LiFTFSI electrolytes taken from Brinkkötter et al.[54] and of EMIMTFSI and EMIMBF$_4$ electrolytes from Gouverneur et al.[23]).

The relative volume flux and mass flux according to eq (6) are shown in **Figure 3**. The relative mass flux (see **Figure 3a**) is negative for most systems, reflecting a net mass flux towards the anode, similar to the findings for neat ILs, see **Figure 2a**. On the contrary, the relative molar volume flux (**Figure 3b**) is zero within error for most systems. This confirms the constant local molar volume as the governing constraint for ion transport even in ternary Li salt in IL electrolytes. We note that with the variation of the Li salt fraction from 0 to 0.4, a large variation of the sample viscosity is covered. The local molar volume constraint is thus valid irrespective of the viscosity of the electrolyte. The only notable exception arises for EMIMBF$_4$/LiBF$_4$-based systems, where the net volume flux appears to be positive. One reason for this could lie in the structure of the BF$_4^-$ anion, different from that of the chemically more closely related anions FSI, TFSI and FTFSI. As during the application of the electric field pulse in eNMR electrolyte degradation reactions occur at the electrodes, differences in the degradation chemistry could influence the net volume flux. If the molar volume of the resulting products strongly differs from that of the initial species, volume could either be created or diminished at the electrodes, leading to a net volume flux through the active volume. This might explain slight deviations of some data points for BF$_4$- and possibly PF$_6$- based samples, see also **Figures 2b and S3**. Indeed, studies of the electrochemical decomposition of BMIMBF$_4$ show that at the anode fluorinated compounds F$_2$, BF$_3$ and B$_2$F$_7^-$ are formed.[55] Despite their high reactivity the gaseous compounds F$_2$ and BF$_3$ might evaporate, causing potential volume loss and thus deviations from the principle of the local volume constraint. In conclusion, apart from the exception of BF$_4^-$, the local volume constraint is shown to be the fundamental conservation law governing the transport processes in ILs and respective electrolytes.

Having proven the local molar volume constraint, we postulate that this constraint does not only apply in eNMR experiments, but in any experimental setup where an incompressible liquid electrolyte is investigated between electrodes under an electric field. Experimental setups are typically open systems concerning momentum exchange, but closed systems concerning volume exchange. In order to compare experimental data to MD simulations, where momentum conservation is an *a priori* assumption, transformations between experimental and simulated data sets have to be performed. Furthermore, for application of theoretical frameworks such as the Onsager formalism, considered in a mass-centered frame with local momentum conservation, reference frame transformations between the lab frame (i.e. volume fixed frame) and the COM frame are required.

Further consequences concern the transference number, which is dependent on the reference frame. **Figure S8** gives Li transference numbers calculated from mobilities as measured in the lab system, i.e. under volume constraint. In agreement with earlier findings they are negative for low to medium Li salt fractions even in the additional systems of this work. This highlights the vehicular Li transport mechanism as a general principle. Only at high Li salt fractions the transference number becomes positive, indicating a cross-over from vehicular to structural Li transport. Red symbols in **Figure S8** show transference numbers after transformation of the mobilities to the COM frame. The values are higher, due to the net mass drift towards the anode, but they remain predominantly negative. A more detailed discussion of transference numbers in different reference frames will be published in a forthcoming theoretical paper.[56] In conclusion, comparison of transference numbers requires information of the reference frame they arise from. We note that in a similar way transference numbers taken under anion blocking conditions systematically differ from those of the COM frame.[11]

As our results show, the laboratory frame, in which electrophoretic mobilities are measured, is identical to a volume-fixed frame, where the net volume flux through any surface perpendicular to the electric field is zero. This implies a *local* volume constraint on a molecular scale. In previous work some of us have argued that the limitation of local dipole fluctuations and mass fluctuations is a constraint which determines ion-ion correlations and explains the magnitude of Onsager coefficients, leading to negative correlations for all coefficients.[11, 25] With the present findings, however, the role of mass fluctuations has to be replaced by limitations of molar volume fluctuations acting as the driving force for the dominance of anticorrelations. The sketch in the Table of contents image illustrates how the displacement of an ion will require a displacement of another ion in the opposite direction in order to achieve a net volume flux of zero. Nevertheless, the general idea of limited local fluctuations remains an important concept to explain the dominance of ion-ion anticorrelations, which dominate in most concentrated ionic systems,[10, 11] while positive correlation coefficients are rare and can only be achieved for strongly coordinating ion species.[28]

More practically, the volume conservation constraint has implications as to the number of experimental parameters required to fully describe the flux of each species and the transference number. In an IL with two ion species and a given total ion conductivity and molar volumes of both species, the local volume constraint leads to a full determination of the ion mobilities, making their measurement obsolete, once the molar volumes of both ion species are known. A ternary (Li salt in IL) system with three ion species is completely determined by two independent quantities, either mobilities or conductivity, if volume conservation is assumed and molar ion volumes of all species are known.

In conclusion, this work provides a deeper understanding of the underlying principles governing transport processes in highly concentrated electrolytes, e.g. IL-based systems. Density measurements are proven to be a very robust technique delivering species-specific size information. The results reveal that local momentum conservation is not fulfilled, it is rather local molar volume conservation, which forms the relevant constraint. Consequently, any volume flux of an ion species in one direction in the electric field is always balanced by a corresponding volume flux of other species to the opposite side. On the molecular scale this implies a negative contribution to the correlation coefficients between the same and different ion species, which explains the dominance of anticorrelations in Onsager descriptions of ion transport. The concept of the volume constraint is largely independent of size or chemical structure, showing that local molar volumes are a central conservation quantity in liquid electrolyte systems. This further implies a non-zero net mass flux, a fact that conflicts with local momentum conservation, generally assumed in MD simulations, requiring reference frame transformations of the latter for compatibility with realistic experimental situations.

**Experimental**

A description of the materials, the electrolyte preparation and the density measurements is found in the SI, sections 1 and 4, respectively. Electrophoretic mobilities $\mu$ of IL cations, Li$^+$ and anions are measured using $^1$H, $^7$Li and $^{19}$F eNMR, respectively. Measures to suppress the influence of thermal convection include using a double stimulated echo pulse sequence (dSTE),[49] see **Figure 1a**, as well as employing a sample holder with Pd electrodes containing a bundle of capillaries. Details of the custom-made setup were published previously,[22] further information on the equipment, experimental parameters and evaluation of phase shifts employing phase-sensitive Lorentz-fitting[27] is given in SI. For each sample, the total conductivity as obtained via mobilities from eNMR is compared to a measurement by impedance spectroscopy, see **Figures S6** and **S7**.

ACKNOWLEDGMENT

We thank Chiara Weiss for assisting with some of the density measurements.

**Supporting Information Available**. A pdf file with additional data, i.e. list of materials employed, description of experimental methods, tabulated mobilities of all species in ILs and ternary electrolytes, densities of ILs and respective electrolytes, procedure and results of calculating molar volumes of all species, phase shift raw data from eNMR, comparison of molar conductivities from eNMR and impedance spectroscopy, and Li transference numbers. This material is available free of charge via the Internet at

http://pubs.acs.org.

Supporting Information for

# Local volume conservation in concentrated electrolytes is governing charge transport in electric fields

Martin Lorenz, Franziska Kilchert, Pinchas Nürnberg, Max Schammer, Arnulf Latz, Birger Horstmann, Monika Schönhoff

**Contents:**

## 1. Materials: Ionic Liquids and Li salts

**Table S1.** Ionic Liquid and Li salt materials and their abbreviations, provider and purity.

| IL | Abbreviation | Purity | Provider |
|---|---|---|---|
| 1-butyl-1-methylpiperidinium bis(trifluoromethylsulfonyl)imide | BmPipTFSI | 99% | IOLITEC |
| 1-butyl-3-methylimidazolium bis(trifluoromethylsulfonyl)imide | BMIMTFSI | 99% | IOLITEC |
| 1-butyl-3-methylimidazolium hexafluorophosphate | BMIMPF$_6$ | 99% | IOLITEC |
| butyltrimethylammonium bis(trifluoromethylsulfonyl)imide | BMATFSI | 99% | IOLITEC |
| ethylmethylimidazolium bis(trifluoromethylsulfonyl)imide | EMIMTFSI | 99% | IOLITEC |
| 1-butyl-1-methylpyrrolidinium bis(trifluoromethylsulfonyl)imide | Pyr$_{14}$TFSI | 99.5% | IOLITEC |
| ethylmethylimidazolium tetrafluoroborate | EMIMBF$_4$ | >98% | IOLITEC |
| 1-methoxyethyl-1-methylpyrrolidinium bis(trifluoromethylsulfonyl)imide | Pyr$_{12O1}$TFSI | 99.9% | SOLVIONIC |
| 1-methoxyethyl-1-methylpyrrolidinium bis(fluorosulfonyl)imide | Pyr$_{12O1}$FSI | 99.9% | SOLVIONIC |
| 1-butyl-1-methylpyrrolidinium bis(fluorosulfonyl)imide | Pyr$_{14}$FSI | 99.9% | SOLVIONIC |
| ethylmethylimidazolium bis(fluorosulfonyl)imide | EMIMFSI | 99.9% | SOLVIONIC |
| Li salt | | | |
| lithium bis(trifluoromethylsulfonyl)imide | LiTFSI | 99% | ALDRICH |
| lithium bis(fluorosulfonyl)imide | LiFSI | >98% | TCI |
| lithium tetrafluoroborate | LiBF$_4$ | 98% | ACROS ORGANICS |
| lithium hexafluorophosphate | LiPF$_6$ | >99.99% | ALDRICH |

All ILs are used without further purification after storage over molecular sieve. The Li salts are dried for at least 8 h at 100°C in high vacuum (10$^{-6}$ hPa). Electrolytes are prepared by mixing appropriate amounts of each IL with the respective lithium salt of the same anion in a glovebox (argon atmosphere, H$_2$O/O$_2$ < 5ppm).

## 2. Experimental parameters and procedures in eNMR

All measurements are conducted on an AVANCE Neo 400 MHz NMR spectrometer (Bruker, Rheinstetten, Germany) equipped with a gradient probe head (DiffBB, Bruker) allowing a maximum magnetic field gradient strength of 17 T·m$^{-1}$. The temperature is adjusted using a heated flow of dry nitrogen, and calibrated with an external thermocouple (PT100, Greisinger electronics) in a 5 mm NMR tube filled with glycol. Electric field pulses are generated by a power source (eNMR - 1000mc, P&L Scientific, Lidingö, Sweden).

In each experiment a series of spectra is taken, while stepwise incrementing the modulus of the voltage $U$, using alternating polarity, up to a maximum value of ~ 100 V for lithium salt in IL electrolytes, or 20-40 V in case of pure IL samples. Parameters fixed in this series, but adapted to the respective sample, are the observation time $\Delta$ (50-100 ms), the gradient pulse duration $\delta$ (2-3 ms) and the gradient strength $g$ (0.5 - 9 T/m). The phase shift $\varphi - \varphi_0$ for each spectrum is determined using Lorentzian-type fitting of the respective spectral line. By increasing $U$ in a series of spectra, the electrophoretic mobility $\mu$ is extracted from the slope in a linear plot of $\Phi - \Phi_0$ against $U$ according to

$$\Phi - \Phi_0 = \gamma \cdot g \cdot \delta \cdot \Delta \cdot \frac{U}{d} \cdot \mu. \tag{S1}$$

The mobilities of all species are measured for at least three different sample preparations of the respective IL or electrolyte and the resulting values are averaged. Resulting errors are the sum of fitting errors, statistical errors and an additional error contribution estimated to 5% to account for other error sources.

## 3. Electrophoretic NMR data for binary systems (pure ILs, two ionic species)

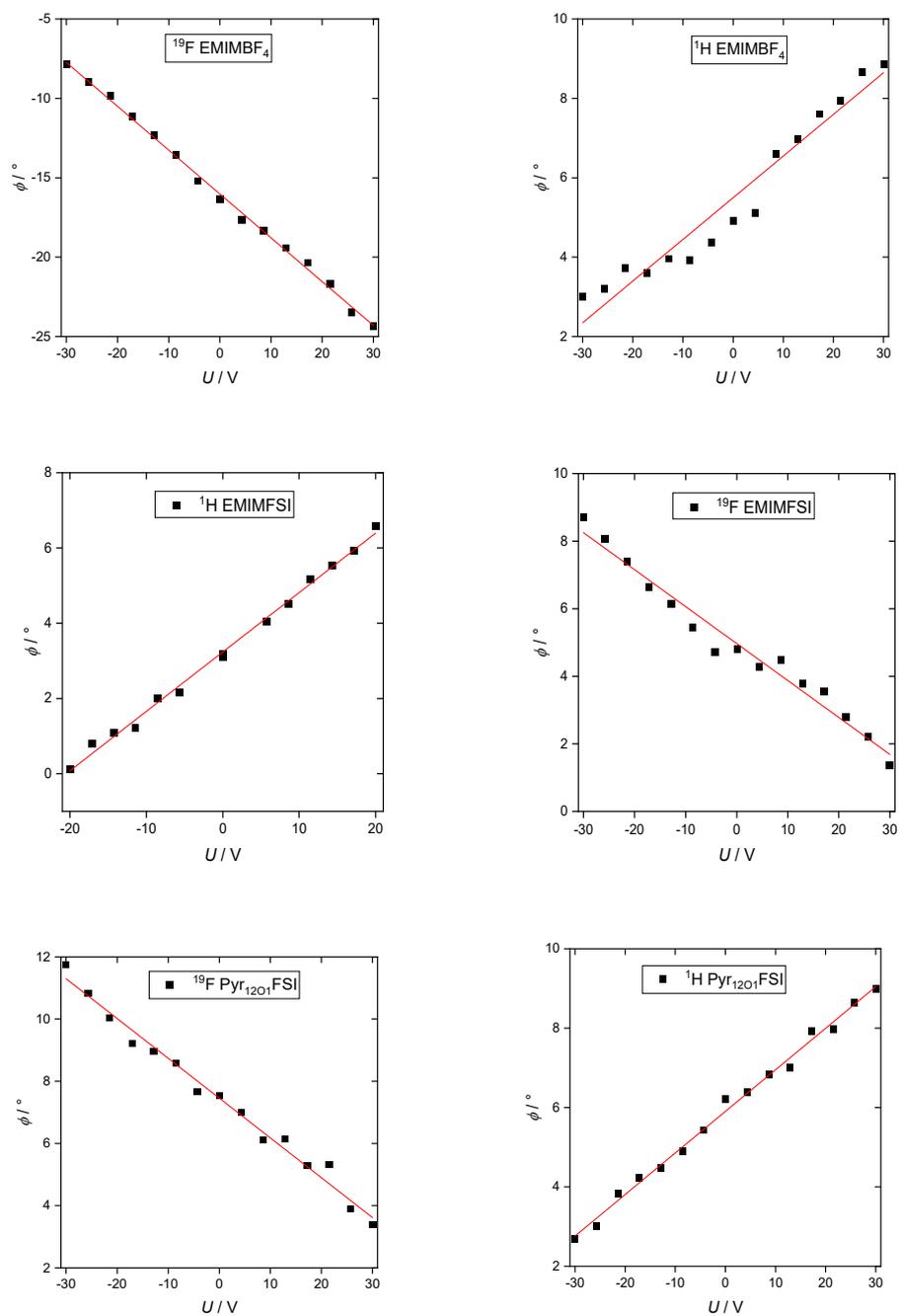

**Figure S1**. Some exemplary phase shift data of selected IL samples. The line represents a linear fit according to equ. (S1).

**Table S2.** Cation and anion mobilities in pure ionic liquids.

| IL | $T$ / K | $\mu_+$ / $(10^{-10}\ m^2V^{-1}s^{-1})$ | $\mu_-$ / $(10^{-10}\ m^2V^{-1}s^{-1})$ |
|---|---|---|---|
| EMIMBF$_4$ | 295 | 6 ± 2 | -13 ± 2 |
| EMIMFSI | 295 | 15 ± 5 | -17 ± 7 |
| Pyr$_{12O1}$FSI | 295 | 4.1 ± 0.7 | -9 ± 2 |
| BMIMPF$_6$ | 295 | 0.7 ± 0.2 | -1.3 ± 0.2 |
| EMIMTFSI | 300 | 11 ± 3 | -10 ± 4 |
| BMATFSI | 300 | 2.6 ± 0.5 | -2.9 ± 0.5 |
| Pyr$_{14}$TFSI | 300 | 4 ± 1 | -6 ± 3 |

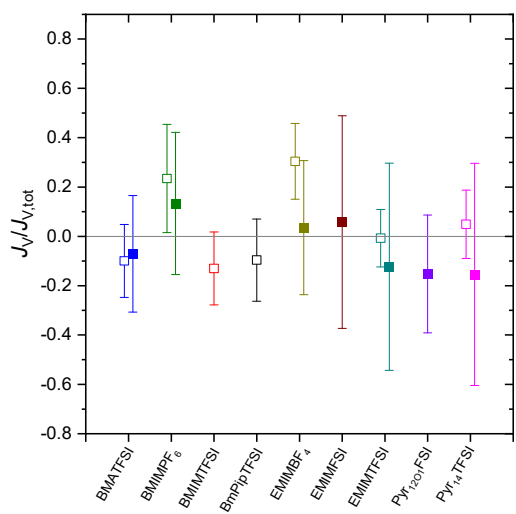

**Figure S2** Comparison of molar volume flux ratios $\frac{J_V}{J_{V,tot}}$ for all investigated ILs using eNMR-data from Gouverneur et al.[1] (open symbols) and of this work (solid symbols).

## 4. Determination of molar volumes of ion species from density measurements

The required partial molar volumes $v_+$ and $v_-$ for the cations and anions in an ionic liquid can be derived from density measurements under variation of the electrolyte composition.

Density measurements of all ILs and prepared electrolytes are performed utilizing a DDM 2910 Automatic Density Meter (Rudolph Research Analytical). Purging the instrument with a flow of dry nitrogen before use ensured the absence of water and oxygen.

The total molar volume $V$ is derived from the molar mass $M$ and the density $\rho$ according to

$$V = \frac{M}{\rho}. \tag{S2}$$

We use an ionic liquid and add varying fractions of the corresponding Li salt with the same anion. By measuring the densities of these electrolytes, information about $V$ in dependence of the lithium salt fraction is accessible. We introduce the molar fractions $x_{\text{Li salt}}$ and $x_{\text{IL}}$ with $x_{\text{Li salt}} + x_{\text{IL}} = 1$. Then, the total molar volume of the mixture, referring to 1 mol of anions is given as

$$V = V_{\text{Li salt}} \cdot x_{\text{Li salt}} + V_{\text{IL}} \cdot x_{\text{IL}} = (V_{\text{Li salt}} - V_{\text{IL}}) \cdot x_{\text{Li salt}} + V_{\text{IL}} \tag{S3}$$

Exemplary plots of the molar volume $V$ for EMIM-based systems with different amounts of respective Li salt are given in Figure S3, where a linear fit indicates agreement with the linearity of eq (S3).

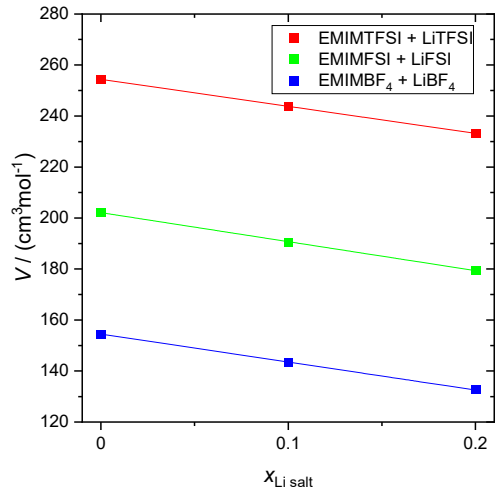

**Figure S3.** Molar volumes $V$ of EMIM-based electrolytes in dependence of the molar lithium salt fraction $x_{\text{Li salt}}$ at 295 K. The straight lines are linear fits according to eq (S3).

This ideal linear relationship also holds true for all other investigated ILs and their corresponding lithium salt in IL electrolytes, as can be seen from the respective graphs in Figure S4. Consequently, in the investigated concentration range the molar volumes $v_+$ and $v_-$ of the ionic species can be considered as independent of the composition. This enables their evaluation from the fits of eq (S3), obtaining the molar volumes of the IL, $V_{\text{IL}}$, and the lithium salt, $V_{\text{Li salt}}$, from the slope and the ordinate intercept, respectively, such that $v_+$ and $v_-$ can be determined according to

$$v_- = V_{\text{Li salt}} - v_{\text{Li}^+} \tag{S4}$$

$$v_+ = V_{\text{IL}} - v_- \tag{S5}$$

The molar lithium volume $v_{\text{Li}^+}$ is given by the effective ionic radius $r_{\text{Li}^+}$, which is well described in literature. For example, Shannon [2] reported a value of $r_{\text{Li}^+} = 76$ pm as the effective ionic radius of a sixfold coordinated lithium ion. Since a coordination number of 6 is a common feature in highly concentrated lithium containing electrolytes, this radius is used in the following for all further evaluations, resulting in a Li$^+$ ion molar volume of

$$v_{\text{Li}^+} = \frac{4}{3} \cdot \pi \cdot N_A \cdot r_{\text{Li}^+} = 1.11 \cdot 10^{-6} \frac{\text{m}^3}{\text{mol}}. \tag{S6}$$

Partial molar volumes of the IL cation and anion obtained from eq (S4) and (S5), are given in Table S3. Details including the underlying density values and molar volumes are given in Table S4, S5 and S6.

**Table S3.** Molar volumes of anions and cations for different ILs, determined at 295 K. Errors are estimated to 3%.

| IL | $v_+$ [cm$^3$/mol] | $v_-$ [cm$^3$/mol] |
|---|---|---|
| EMIMTFSI | 107 ± 3 | 147 ± 4 |
| EMIMBF$_4$ | 111 ± 3 | 44 ± 1 |
| Pyr$_{14}$TFSI | 156 ± 5 | 146 ± 4 |
| BmPipTFSI | 166 ± 5 | 149 ± 4 |
| BMATFSI | 140 ± 4 | 144 ± 4 |
| BMIMTFSI | 145 ± 4 | 146 ± 4 |
| BMIMPF$_6$ | 144 ± 4 | 63 ± 2 |
| EMIMFSI | 115 ± 3 | 87 ± 3 |
| Pyr$_{12O1}$FSI | 145 ± 4 | 91 ± 3 |
| Pyr$_{12O1}$TFSI | 146 ± 4 | 145 ± 4 |

The variety of ILs examined makes it possible to compare the molar volume of an ionic species for different counterions. The molar volumes seem to be rather independent of the respective counterion, which confirms the assumption of largely composition independent molar ion volumes. TFSI molar volumes differ by only up to 3 % depending on the cation. EMIM molar volumes show a larger range of variation, which is possibly an effect of an optimization of the packing in combination with the large variation of anion sizes employed.

Electrophoretic mobilities in this work are determined at room temperature (either at T = 295 K or at 300 K). Density data acquired at 295 K are given in Table S4 and the resulting molar volumes in Table S3. In addition, densities are determined at 300 K, see Table S5, yielding the molar volumes in Table S6. In all further calculations electrophoretic mobilities are combined with molar volumes acquired at the same temperature. In some cases the temperatures employed in eNMR experiments from literature lie within the range from 295 K to 300 K. As differences of the molar volumes in this small temperature range are negligible (compare Tables (S3) and (S6)), mobilities from literature are always combined with the molar volumes of the closest temperature, i.e. either from Table (S3) or Table (S6), respectively.

**Table S4.** Densities of ILs and ternary electrolytes employed for the determination of molar volumes at 295 K.

| $x_{Li\text{-salt}}$ | 0 | 0.1 | 0.2 |
|---|---|---|---|
| | ρ / kg·m$^3$ | | |
| EMIMTFSI | 1538.2 | 1562.2 | 1588.7 |
| EMIMBF$_4$ | 1280.8 | 1308.4 | 1335.7 |
| Pyr$_{14}$TFSI | 1397.6 | 1425.9 | 1457.8 |
| BmPipTFSI | 1384.2 | 1412 | 1440.6 |
| BMATFSI | 1395.3 | 1426.6 | 1460.9 |
| BMIMTFSI | 1439.6 | 1466.5 | 1496.7 |
| BMIMPF$_6$ | 1370.9 | 1403.9 | 1442.4 |
| EMIMFSI | 1440.2 | 1474.9 | 1506.6 |
| Pyr$_{12O1}$FSI | 1376.5 | 1404.1 | 1434.7 |
| Pyr$_{12O1}$TFSI | 1456.6 | 1483.1 | 1512.8 |

**Table S5.** Densities of ILs and ternary electrolytes employed for the determination of molar volumes at 300 K.

| $x_{Li-salt}$ | 0 | 0.1 | 0.143 | 0.2 | 0.3 | 0.4 |
|---|---|---|---|---|---|---|
| | $\rho$ / kg·m$^3$ | | | | | |
| EMIMTFSI | 1534.5 | | | 1584.6 | 1610.1 | |
| EMIMBF$_4$ | 1278.8 | 1303.3 | | 1332.9 | | |
| Pyr$_{14}$TFSI | 1393.2 | 1421.4 | | 1453.6 | | |
| Pyr$_{14}$FSI | 1311 | 1334.8 | 1347.9 | | | |
| BmPipTFSI | 1379.6 | 1407.9 | | 1440.6 | | |
| BMATFSI | 1391.1 | 1424 | | 1458.3 | | |
| BMIMTFSI | 1435 | 1461.8 | | 1491.7 | | |
| BMIMPF$_6$ | 1366.9 | 1400.2 | | 1423.3 | | |
| EMIMFSI | 1459.9 | | | 1511.8 | 1544.1 | |
| Pyr$_{12O1}$FSI | 1369.2 | | | 1431.8 | 1467.9 | 1511.7 |
| Pyr$_{12O1}$TFSI | 1451.2 | 1477.7 | | 1507.5 | 1540.7 | 1578.1 |

**Table S6.** Molar volumes of anions and cations for different ILs, measured at 300 K. Errors are estimated as 3 %.

| IL | $v_+$ [cm$^3$/mol] | $v_-$ [cm$^3$/mol] |
|---|---|---|
| EMIMTFSI | 106 ± 3 | 149 ± 4 |
| EMIMBF$_4$ | 111 ± 3 | 44 ± 1 |
| Pyr$_{14}$TFSI | 157 ± 5 | 146 ± 4 |
| Pyr$_{14}$FSI | 148 ± 4 | 98 ± 3 |
| BmPipTFSI | 172 ± 5 | 145 ± 4 |
| BMATFSI | 142 ± 4 | 143 ± 4 |
| BMIMTFSI | 145 ± 4 | 147 ± 4 |
| BMIMPF$_6$ | 135 ± 4 | 73 ± 2 |
| EMIMFSI | 105 ± 3 | 95 ± 3 |
| Pyr$_{12O1}$FSI | 148 ± 4 | 89 ± 3 |
| Pyr$_{12O1}$TFSI | 147 ± 4 | 146 ± 4 |

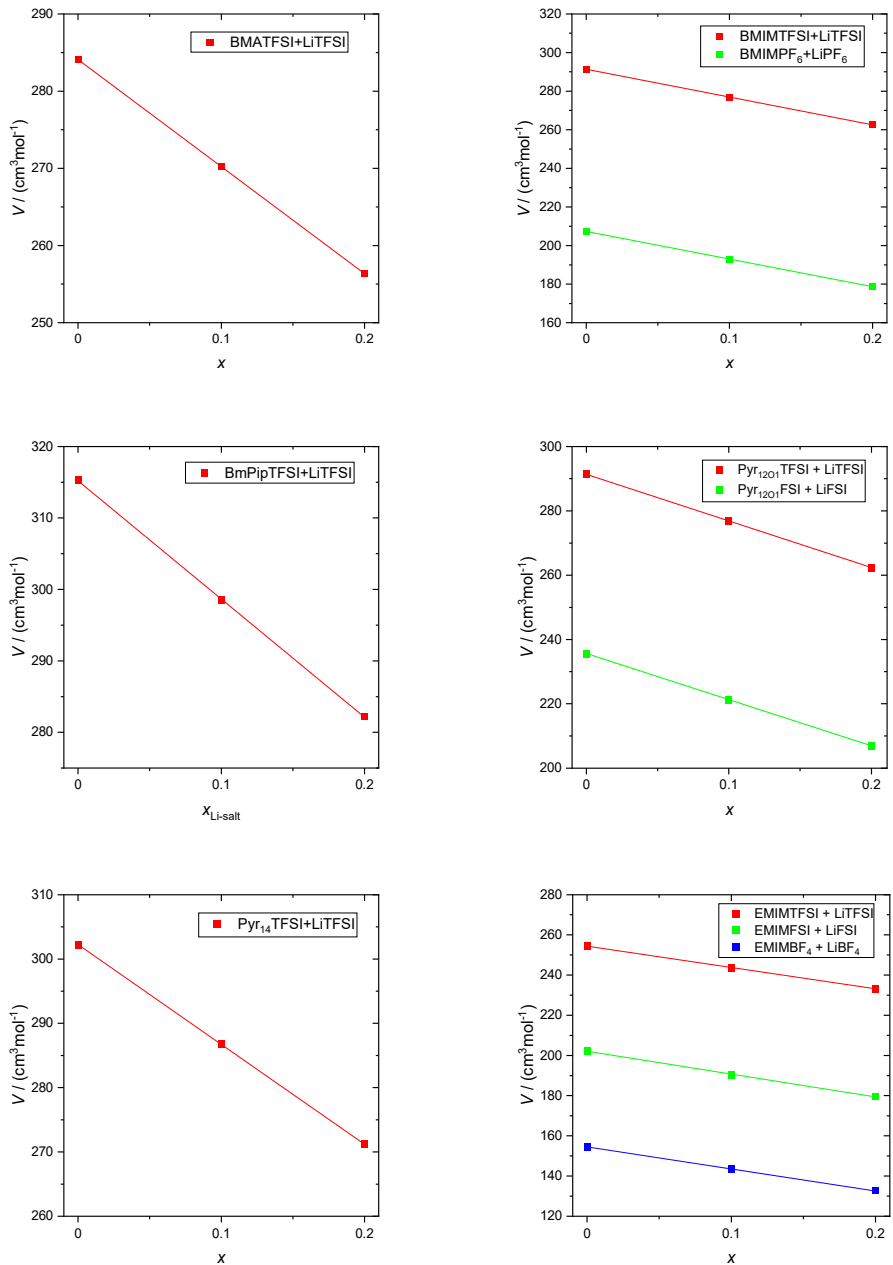

**Figure S4.** Molar volumes $V$ obtained from density measurements for $(IL)_{1-x}(Li\ salt)_x$ electrolytes in dependence of the molar lithium salt fraction $x$ at 295 K. Lines are linear fits employed to determine $V_{IL}$ and $V_{Li\ salt}$ according to eq (S3).

## 5. Electrophoretic NMR data for ternary systems (Li salt in IL, three ionic species)

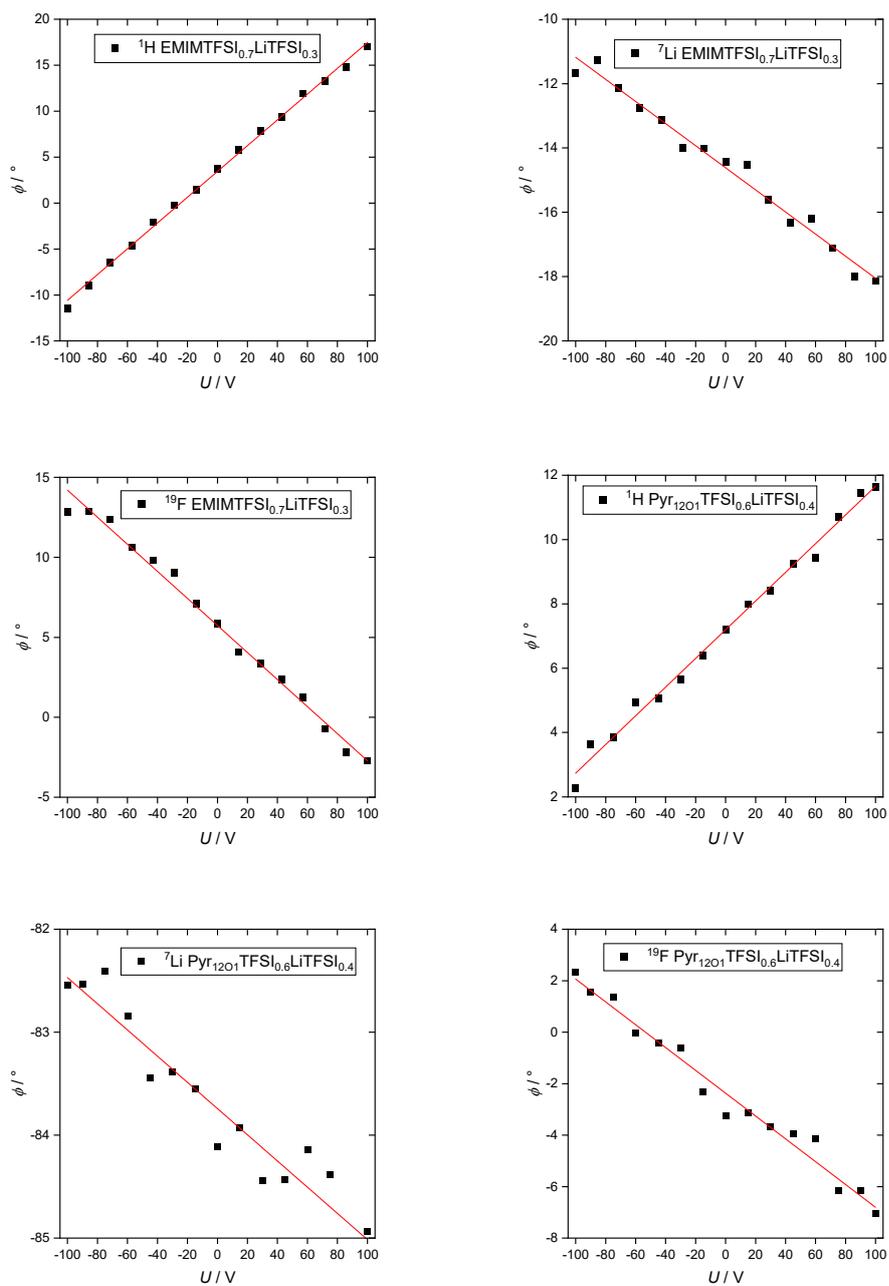

**Figure S5**. Some exemplary phase shift data of selected ternary samples of Li salt in IL. The line represents a linear fit according to eq (S1).

**Table S7.** Electrophoretic mobilities of IL cation, lithium and common anion in Li salt in Il electrolytes at 300 K. The values are averaged over at least three independent experiments.

| Electrolyte | $\mu_+$ / $10^{-10}$ m$^2$V$^{-1}$s$^{-1}$ | $\mu_-$ / $10^{-10}$ m$^2$V$^{-1}$s$^{-1}$ | $\mu_{Li}$ / $10^{-10}$ m$^2$V$^{-1}$s$^{-1}$ |
|---|---|---|---|
| EMIMFSI$_{0.9}$LiFSI$_{0.1}$ | 17 ± 1 | -12 ± 2 | -5 ± 1 |
| EMIMFSI$_{0.86}$LiFSI$_{0.14}$ | 13.1 ± 0.3 | -12 ± 3 | -3 ± 1 |
| EMIMFSI$_{0.7}$LiFSI$_{0.3}$ | 7 ± 2 | -8 ± 2 | -3 ± 2 |
| EMIMTFSI$_{0.9}$LiTFSI$_{0.1}$ | 12 ± 2 | -4.8 ± 0.9 | -5 ± 2 |
| EMIMTFSI$_{0.86}$LiTFSI$_{0.14}$ | 8 ± 1 | -4 ± 2 | -5.7 ± 0.8 |
| EMIMTFSI$_{0.7}$LiTFSI$_{0.3}$ | 4.1 ± 0.5 | -2.1 ± 0.3 | -1.7 ± 0.3 |
| Pyr$_{12O1}$FSI$_{0.6}$LiFSI$_{0.4}$ | 1.9 ± 0.3 | -2.0 ± 0.2 | 0.3 ± 0.1 |
| Pyr$_{12O1}$TFSI$_{0.6}$LiTFSI$_{0.4}$ | 0.52 ± 0.09 | -0.32 ± 0.07 | -0.20 + 0.04 |
| (EMIMBF$_4$)$_{0.86}$(LiBF$_4$)$_{0.14}$ | 6 ± 2 | -7 ± 2 | -7 ± 1 |
| Pyr$_{14}$FSI$_{0.9}$LiFSI$_{0.1}$ | 4.6 ± 0.9 | -7 ± 2 | -3.9 ± 0.5 |
| Pyr$_{14}$FSI$_{0.86}$LiFSI$_{0.14}$ | 4.3 ± 0.6 | -7 ± 1 | -2.9 ± 0.9 |
| Pyr$_{14}$TFSI$_{0.9}$LiTFSI$_{0.1}$ | 2.6 ± 0.7 | -2.7 ± 0.5 | -1.7 ± 0.3 |
| Pyr$_{14}$TFSI$_{0.86}$LiTFSI$_{0.14}$ | 2.01 ± 0.06 | -2.0 ± 0.2 | -1.2 ± 0.2 |

## 6. Comparison of conductivity values from eNMR and IS

As a quality check of the results, especially with respect to potential artefacts like convection and electroosmosis, molar conductivities measured by means of impedance spectroscopy are compared with the molar conductivity calculated from eNMR mobilities of all species in the system. The latter can be obtained according to

$$\Lambda_{\text{eNMR}} = z \cdot F \cdot (\mu_+ - \mu_-) \quad (S7)$$

for the case of a pure IL only including one cationic and anionic species each, where $F$ is the Faraday constant and $z$ is the valency.[1] Note that $\mu_-$ is negative due to the opposite drift direction of the anions. In case of a lithium salt in IL electrolyte with a Li salt fraction $x$ the contributions of three species with their respective valences $z_i$ are considered in

$$\Lambda_{\text{eNMR}} = F \cdot \left( (1-x) \cdot z_+ \cdot \mu_+ + z_- \cdot \mu_- + x \cdot z_{\text{Li}} \cdot \mu_{\text{Li}} \right) \quad (S8)$$

Conductivity measurements by impedance spectroscopy (IS) are carried out using an Alpha-S frequency analyzer (Novocontrol) in combination with a Microcell HC measuring stand (rhd instruments). The sample is placed inside a hermetically sealed platinum crucible and the real and imaginary part of the complex impedance are determined in the frequency range between $10^0$-$10^7$ Hz with a voltage amplitude of 10 mV. The same applied to the measurement of 0.01 M KCl in water, which is used as a standard for the determination of the cell constant. The direct current conductivity of interest, $\sigma_{\text{dc}}$, could be obtained at the minimum of the imaginary part in a Nyquist plot of the frequency dependent data. The molar conductivity $\Lambda_{\text{dc}}$ can be calculated as

$$\Lambda_{\text{dc}} = \frac{\sigma_{\text{dc}} \cdot M}{\rho} \quad (S9)$$

from the conductivity $\sigma_{\text{dc}}$ as obtained from impedance measurements, with the density $\rho$ being known. Comparisons of $\Lambda_{\text{eNMR}}$ and $\Lambda_{\text{dc}}$ for the ILs and ternary systems measured in this work are shown in Figures S6 and S7, respectively. There is generally an excellent agreement between both datasets.

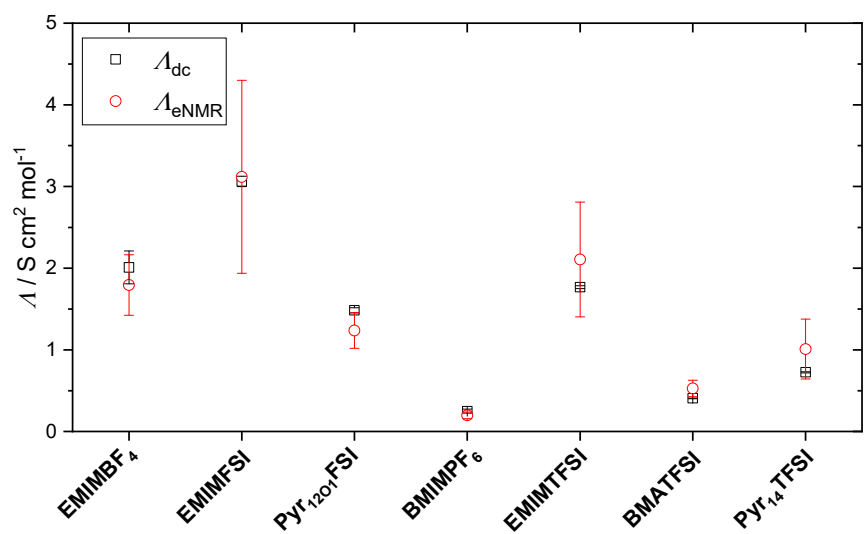

**Figure S6.** Comparison of molar conductivities determined by means of impedance spectroscopy (black triangles) and calculated from eNMR mobilities (red circles) for ILs. For electrophoretic mobilities published by Gouverneur et al.[1] we refer to the corresponding confirmation of agreement of $\Lambda_{\mathrm{eNMR}}$ and $\Lambda_{\mathrm{dc}}$ in the respective publication.

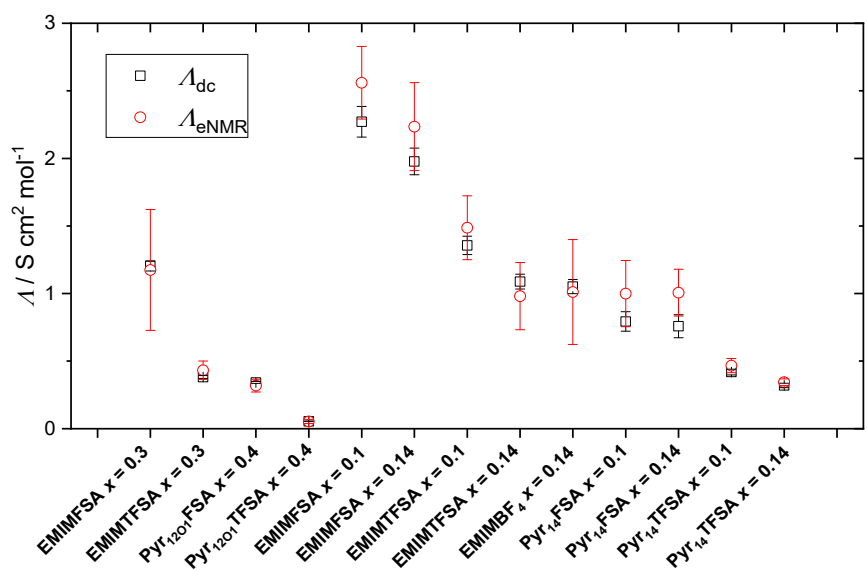

**Figure S7.** Comparison of molar conductivities determined by means of impedance spectroscopy (black squares) and calculated from eNMR mobilities (red circles) for ternary electrolytes with different molar fractions $x$ of the respective Li salt. For electrophoretic mobilities published by Brinkkötter et al.[3] and Gouverneur et al.[4] we refer to the corresponding confirmation of agreement of $\Lambda_{\mathrm{eNMR}}$ and $\Lambda_{\mathrm{dc}}$ in the respective publication.

## 7. Li transference numbers

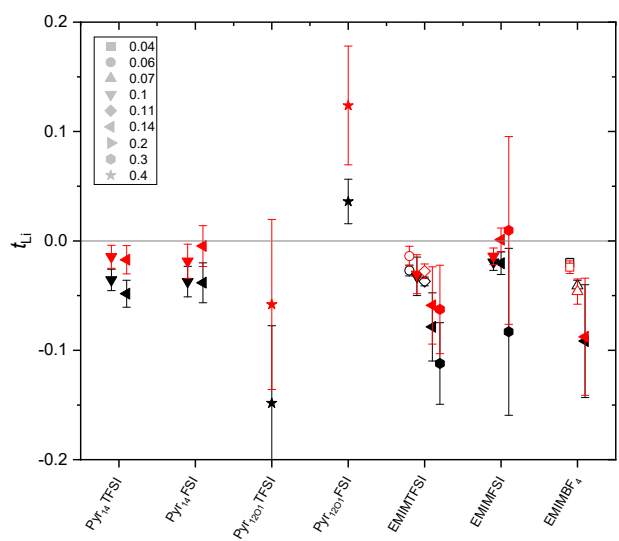

**Figure S8**. Transference numbers of the Li salt in IL electrolytes studied in this work (full symbols) and from Gouverneur et al.[3] (open symbols), calculated from electrophoretic mobilities in the volume-fixed frame (i.e. lab frame; black symbols) and transformed to the COM frame (red symbols). The legend indicates the molar fraction of Li salt.